# Neutralization of slow helium ions scattered from single crystalline aluminum and tantalum surfaces and their oxides

Barbara Bruckner[a,b,*], Peter Bauer[b] and Daniel Primetzhofer[a]

[a] Department of Physics and Astronomy, Uppsala University, Box 516, S-751 20 Uppsala, Sweden

[b] Johannes-Kepler Universität Linz, IEP-AOP, Altenbergerstraße 69, A-4040 Linz, Austria

*barbara.bruckner@physics.uu.se

**Abstract**

We investigated the impact of surface oxygen on the ion yield for He$^+$ ions scattered from different single crystalline surfaces in low-energy ion scattering. Initially clean Al(111) and Ta(111) were exposed to molecular oxygen and ion spectra for different oxidation stages and different primary energies were recorded. A comparison of ion yields normalized to the differential scattering cross section as well as experimental factors allows obtaining information about the influence of oxygen on charge exchange processes. The decrease in the ion yield of both metals with exposure cannot be explained by different surface coverages exclusively, but requires the neutralization efficiency to be dependent on the chemical structure of the surface. For Ta, additionally, a different energy dependency of the ion yield obtained in the metal and oxide occurs. The ion yield for O shows in both surfaces a significantly weaker energy dependency than the investigated metals.

# 1. Introduction

The formation of metal oxides on surfaces is of high technological interest due to their relevance in e.g. catalysis and corrosion processes or as active component in sensors. The initial stages of oxidation processes are often studied via scanning tunneling microscopy and spectroscopy [1]. Also X-ray photoelectron spectroscopy (XPS) [2] and low electron energy diffraction (LEED) are widely used. A direct and non-destructive alternative to study composition and structure of surfaces simultaneously with sub-nanometer resolution is low energy ion scattering (LEIS). Typical applications of LEIS are found in surface structure analysis [3], e.g. for catalytic surfaces [4], monitoring of in-situ growth [5,6] or for real space crystallography of the surface [7]. Ultimate surface sensitivity results from the high neutralization probabilities of the commonly employed noble gas ions in an energy regime with primary energies $E_0 \leq 5$ keV. When single crystals are used as targets, ions scattered from deeper layers can be additionally suppressed due to shadowing and blocking effects, yielding even enhanced surface sensitivity for noble gas ions [8] and excellent sensitivity also for alkali projectiles [9,10].

The ion fraction $P^+$ is the physical quantity, which describes the integral charge exchange behavior of backscattered projectiles. The two main charge exchange mechanism in LEIS using noble gas projectiles are (a) Auger neutralization (AN) and (b) resonant processes – ionization (RI) and neutralization (RN) [11]. Auger neutralization along the ion trajectory is possible for all primary energies and is considered to depend exclusively on the interaction time of the projectile and the target. If only Auger process are possible $P^+$ scales exponentially with the inverse perpendicular velocity of the projectile $v_\perp$ and a characteristic velocity $v_c$ [12]:

$$P^+ = \exp(-v_c/v_\perp) \qquad \text{Eq. (1)}$$

For the resonant processes to occur, the projectile-nuclei distance in the collision has to be smaller than a critical distance $r_c$, which is projectile-target dependent. For a given scattering geometry, this minimum distance can be converted in a threshold energy $E_{th}$, above which the resonant charge exchange processes (RN and RI) are enabled. In this resonant regime, projectiles scattered from deeper layers and re-ionized in a final close collision near the surface can additionally contribute to the spectra of backscattered ions [11].

The experimentally accessible quantity in typical LEIS analysis is the ion yield $Y^+$. This quantity, in turn, is trivially dependent on the ion fraction and the surface coverage. In the ideal case, $P^+$ is known, the ion yield is proportional to the surface coverage, and LEIS can be used to quantify surface concentrations. However, for a series of projectile-target systems matrix effects have been observed [13,14]. To obtain a better understanding of the possible charge exchange mechanisms, independent information on the ion fraction and on the surface coverage is necessary. Thus, one needs to investigate samples with well-known surface structures and composition, therefore, typically single crystals are used allowing also the suppression of contributions of deeper layers via channeling and blocking effects [8]. In general, the information depth in LEIS increases with $E > E_{th}$, which can cause additional systematic uncertainties in quantification. For several systems this influence from sub-surface contributions has been assessed, in particular for Au, Al and Cu [8,15,16], which indicated a dependence on surface orientation, with smaller contributions for less open surfaces. For the latter the sub-surface signal was commonly found small compared to the signal from the outermost layer. Finally, also impurities can complicate the surface structure and hamper direct evaluation of ion fractions. In particular, O significantly influences the intensity and shape of spectra obtained from backscattered ions, and matrix effects have been observed for several different systems [6,13,14].

In this contribution, we study the influence of surface oxygen on energy dependence and intensity of the ion yields for single crystalline Al(111) and Ta(111) surfaces. Specifically, we have exposed initially clean surfaces to different doses of oxygen and continually recorded the ion signal obtained, both from the metal constituent and from oxygen. This procedure was performed for a number of different ion energies. The choice of Al and Ta is made due to their similarly low threshold energy of $E_{th} \sim 300$ eV [17] and different surface geometry (fcc vs. bcc).

## 2. Methods

For the present investigations we used the electrostatic analyzer (ESA)-LEIS setup Minimobis [18] at the Johannes Kepler University Linz. The scattering chamber features a base pressure of $< 5 \times 10^{-10}$ mbar. Additionally, $LN_2$ cooling traps are used to optimize the base pressure during experiments. The measurement geometry is fixed: the primary beam hits the target in normal incidence and only ions leaving the target in a direction equivalent to a scattering angle of $\vartheta = 136° \pm 1°$ are detected by micro channel plates, with an azimuthal acceptance angle of $2\pi$.

The yield of backscattered ions from species $i$, $Y_i^+$, depends on its the surface coverage $c_i$ in atoms/cm², the ion fraction, the differential scattering cross section $(d\sigma/d\Omega)_i$, the number of primary ions $N_0$ as well as on setup specific factors: the spectrometer efficiency $\eta_i^+$, the transmission function of the cylindrical mirror analyzer $T(E)$ and the detector solid angle element $d\Omega$.

$$Y_i^+ = c_i \cdot P_i^+ \cdot N_0 \cdot \left(\frac{d\sigma}{d\Omega}\right)_i \cdot \eta_i^+ \cdot T(E) \cdot d\Omega \qquad \text{Eq. (2)}$$

The beam current of typically several hundred pA, which provides $N_0$, is measured with the use of a low suppression voltage to avoid contributions from secondary electrons. The low sputter rate for He in normal incidence [19] and the low primary ion doses below $10^{15}$ at/cm² permits to neglect any sputtering effects from the primary beam during measurements.

For the calculation of the scattering cross sections, in the LEIS regime, screened Coulomb potentials are used. We chose the widely applied Ziegler-Biersack-Littmark (ZBL) or universal potential [20] with no additional screening length correction. Note, that uncertainties in the employed potential will systematically influence absolute evaluation of either surface coverage or ion fraction from ion yields.

The transmission function of the cylindrical mirror analyzer is obtained from measurements performed on a clean polycrystalline Cu sample – a well-known system [21]. In general, $T(E)$ can be expressed as a product of an energy dependent part, which is, in good approximation, expected to be proportional to $E$, and a scaling factor $j$ $T(E) = j \cdot E$.

Before measurements, all investigated samples were cleaned by 3 keV $Ar^+$ sputter and annealing cycles to remove surface contaminations. This cleaning procedure was performed until no increase in the current-normalized surface peak area of the signal of the metal was visible. For the exposure measurements molecular oxygen was introduced into the scattering chamber with maximum pressures up to $10^{-7}$ mbar. During exposures the ion yields of He for scattering on oxygen and the host material are recorded for constant primary energy. This procedure is performed until a saturation in the surface area of oxygen and the host material is achieved. A stable surface oxide is reached at exposure doses of ~ 1200 L for Al [2] and ~ 10 L for Ta [22,23] with 1 L = $1.33 \times 10^{-6}$ mbar·s. Several series were performed for different primary ion energies ranging from 0.65 keV to 3 keV. All the presented measurements were conducted at room temperature.

## 3. Results and discussion

### 3.1 Peak shape analysis

Figure 1 shows the ion signal for 3 keV $He^+$ scattered from a clean, single-crystalline Al(111) surface. Neglecting electronic energy loss, the final energy of the projectile after a binary collision (peak position of the ion signal) can be expressed in terms of the primary energy $E_0$ and the kinematic factor $k$. The latter depends on the masses of the projectile and target atom as well as the scattering angle. However, the peak position, i.e. the maximum intensity of backscattered projectiles, typically is found at a lower energy than $kE_0$ due to interactions of the projectile with the target electrons causing a deceleration of the projectile (electronic stopping). The asymmetric shape of the Al peak is mainly due to contributions of re-ionized projectiles scattered from the surface [15]. Projectiles scattered from deeper layers, which are re-ionized on the exit trajectory, result in a continuous background for $E_{th} < E < kE_0$. The tail at $E > kE_0$ contains dual and plural scattering contributions.

To obtain the ion yield for Al in Fig. 3, the peak can be fitted with two Gaussians under consideration of the background (dash-dotted line). The peak position of the two Gaussians differs by ~ 20 eV, representing contributions from either projectiles surviving AN (black dashed line) or from re-ionized projectiles (black dotted line), both scattered from the surface [15]. If the distance between the two peaks were larger than their combined FWHM, information on both Auger neutralization and re-ionization efficiency could be straightforward deduced. However, typically these two contributions cannot be disentangled as can be seen in Fig. 1. Alternatively, the integral over the peak for Al can be evaluated again with subtraction of the background. The resulting peak areas for Al obtained with either two Gaussians or the integral – both without contributions from the background – differ by less than 5 %. Due to the inseparability of the contributions from either survivals or re-ionized projectiles, we focus on the integral over the peak.

To improve comparability between the ion yields for different materials and primary energies, we typically plot normalized ion signals with

$$A_i^+ \equiv \frac{Y_i^+}{N_0 \cdot (d\sigma/d\Omega)_i \cdot \eta_i^+ \cdot E \cdot d\Omega} = j \cdot c_i \cdot P_i^+. \qquad \text{Eq. (3)}$$

According to Eq. 2, $A_i^+$ corresponds to the product of surface coverage and ion fraction with a constant scaling factor $j$. Consequently, information about the neutralization efficiency can be obtained from these normalized signals.

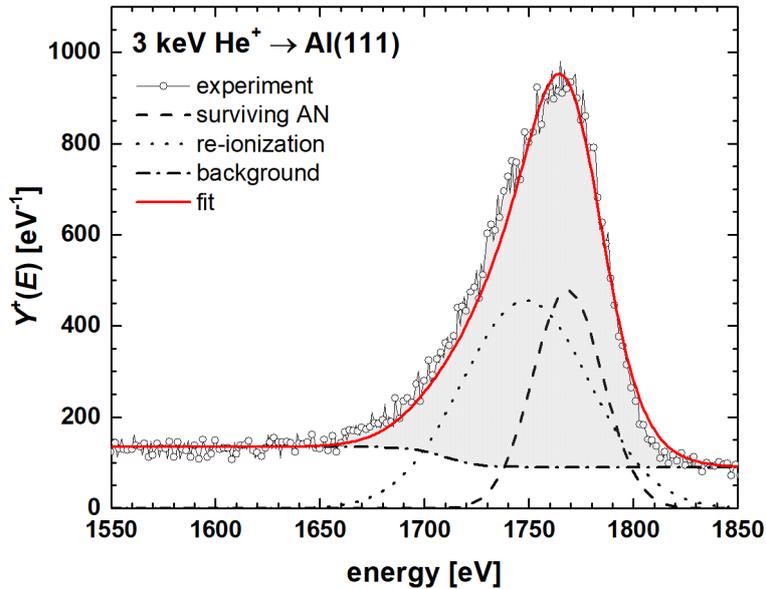

Fig. 1: Experimental energy spectrum of 3 keV He$^+$ ions scattered from an Al(111) surface. The red solid curves corresponds to a fit consisting of two Gaussians (dashed and dotted lines) with a difference of ~ 20 eV in the peak position under consideration of the background (dash-dotted line). The grey colored area corresponds to the integral of the Al peak without contributions from the background.

3.2 Al(111) surface

Figure 2 shows the energy spectra of He$^+$ ions backscattered from either a clean Al(111) or an oxidized Al(111) surface ($\geq$ 500 L) with a primary energy of (a) 1 keV and (b) 3 keV. The spectra obtained from the metal are depicted as black open squares; the red open circles correspond to the ion signals from the oxide. In the latter, the peak area for Al has significantly decreased compared to the metal and a clear peak associated with single scattering from O is visible for both energies. In the following, we will compare intensities of the peak areas of both Al and O as well as the overall shape of the background in the metal and the oxide.

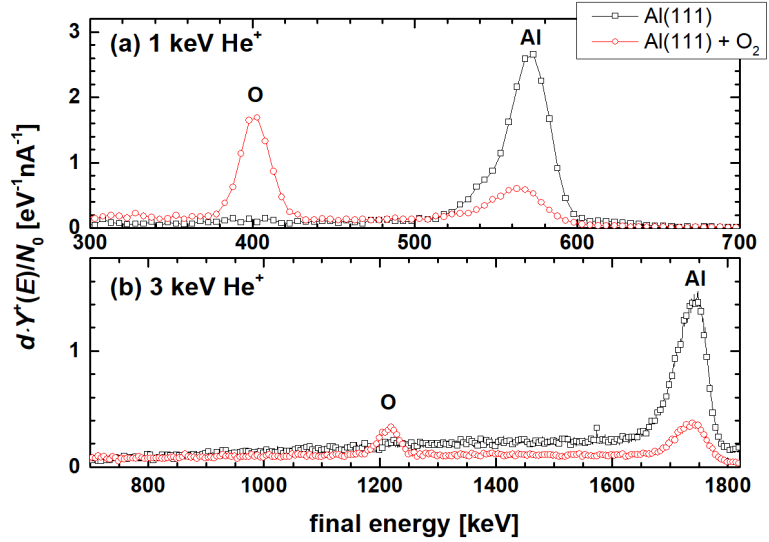

Fig. 2: Experimental ion spectra for (a) 1 keV and (b) 3 keV He$^+$ scattered from a clean Al(111) as well as an oxidized Al(111) surface in black and red, respectively. The yields are normalized to the number of primary ions as well as the setup specific parameters $d$ (detection and the energy dependent transmission efficiency).

For a primary energy of 1 keV, the spectrum recorded for the oxidized Al(111) shows a lower intensity of the Al peak compared to the clean metal with $A^+_{Al,AlO} \sim 0.26 \cdot A^+_{Al,Al}$. The surface coverage for Al in Al(111) and in a stoichiometric Al$_2$O$_3$, however, decreases only by ~ 32 %. This decrease in the surface yield for Al, therefore, can only be explained by less Al atoms in the surface contributing to the ion signal or by a change in the ion fraction. Different experimental and theoretical studies in literature indicate oxygen sitting 0.58 Å – 0.7 Å above the outermost Al(111) layer [24–27]. This structure is indeed expected to result in lower Al signals in the oxide than expected from a planar Al$_2$O$_3$ surface as the effective way of ions in regions with significant electron density would be increased. Nevertheless, according the definition of Eq. 2 this would be a matrix effect, but at the same time exemplifies the difficulties associated with both terms, surface concentration and ion fraction, hampering their separation.

In the spectrum of the oxide a comparison of the ion yields for Al and O – both normalized according to Eq. 3 – yields a significantly smaller signal for Al than for O with $A^+_{Al,AlO} \sim 0.3 \cdot A^+_{O,AlO}$. Note, this normalized yield $A^+_i$ is proportional to the product of ion fraction and surface coverage. The assumption of a stoichiometric Al$_2$O$_3$ surface yields that at 1 keV $P^+_{Al,AlO}$ is ~ 55 % smaller than $P^+_{O,AlO}$.

In the 3 keV spectrum, depicted in Fig. 2(b), the Al yield in the oxide is lower by a factor of ~ 4.3 compared to the clean metal, i.e. $A^+_{Al,AlO} \sim 0.23 \cdot A^+_{Al,Al}$, similarly as for 1 keV. However, a comparison of the normalized O and Al signals, results in similar signal intensities, with $A^+_{Al,AlO} \sim 1.04 \cdot A^+_{O,AlO}$. The large difference in the normalized yields of Al and O at 1 keV and 3 keV can be explained by either a completely different velocity scaling of the ion fractions for O and Al or different information depths, and therefore, different oxygen coverages for the two primary energies, albeit we followed the same sample preparation and oxidation protocol. Additionally, previous investigations showed only minor sub-surface contributions to the ion yields to be expected in the employed energy regime for the investigated system [15].

An overall investigation of the shape and intensity of the ion spectra in Fig. 2 shows, that not only the ion yield of Al differs between the metallic and oxidized surface, but the oxidation can also have an influence on the re-ionization background as it was observed for Zn and Ta [6]. For a primary energy of 1 keV, the background in Fig. 2 at half the primary energy due to projectiles scattered from deeper layers does not change with oxidation while for 3 keV, it is lower in the oxide than in the metal, similarly as for Ta [6]. However, this difference in the background with its low intensity with respect to the single scattering peaks cannot explain the observed different scaling with energy in the ion fraction for O and Al. In the following, we therefore, exclusively focus on the ion yields deduced from the peak signals. For these, also the apparent difference in scattering kinematics is expected to be only of minor importance [28].

In Fig. 3 the ion yields for Al and O as a function of exposure are plotted for different primary energies: 0.65 keV (green asterisks), 1 keV (black squares), 1.5 keV (red circles) and 3 keV (blue triangles). Panel (a) depicts the ion yield $A^+_{Al,AlO}$ normalized according to Eq. 3. The signal intensity at a given exposure, i.e. surface composition, is found to vary with primary energy, which due to the definition of $A^+_i$, can be either due to different ion fractions or different information depths. However, the information depth is supposed to have only a minor influence on the ion yield [15,16] and the occurring differences can be exclusively explained by the expected energy dependence of the ion fraction for Al [15]. In panel (b) we, therefore, present ion yields normalized to the initial value of the exposure curves. The general behavior of the curves, such as the intensities obtained for clean and oxidized surfaces is found virtually independent of the primary ion energy. For all investigated energies, the ion signal of Al stays constant until an exposure of several L indicating a slow adsorption of oxygen on the clean Al(111) surface [1]. With ongoing exposure the ion signal of Al decreases until the signal reaches a steady plateau between several hundred L and stays constant up to ~ 1000 L. Further exposure did not yield any differences in the Al yields. According to an XPS study [2] the surface oxide formed during the exposure is found to have a thickness of ~ 2.4 Å for doses of ~ 1200 L.

The corresponding ion yields of the oxygen peak $A^+_{O,AlO}$ as a function of exposure are displayed in Fig. 3(c). The comparable slow increase in the oxygen signal matches the decrease of the Al signal, indicating a low sticking probability of oxygen on clean Al(111). In literature, sticking coefficients of $s_O = 0.005$ are reported for early oxidation [1]. The increase in the O surface concentration with exposure dose can be expressed in terms of standard statistical growth models: $c_i \propto [1 \pm \exp(-L/L_0)]$ with the exposure dose $L$ and a characteristic exposure parameter $L_0$. We included a fit for the primary energy of 1 keV in Fig. 3(b) and (c) with $L_0 = 100$ L. The fit quality illustrates, that the ion yields for both Al and O can well be described in a single model for the oxidation process.

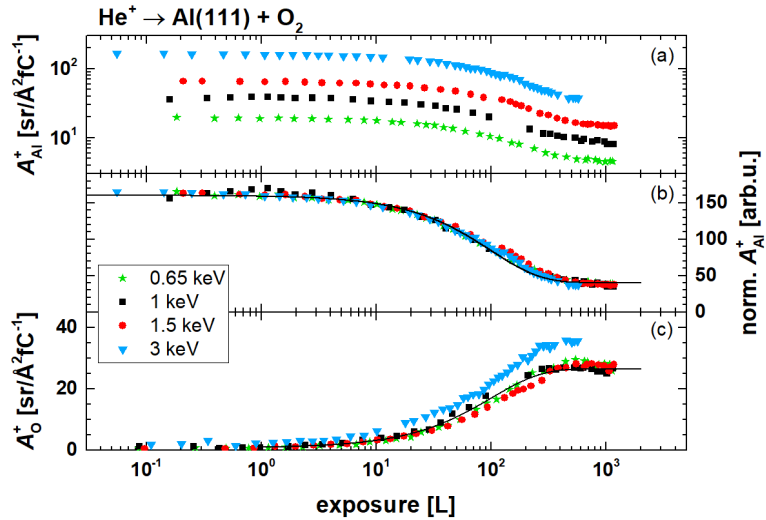

Fig. 3: Ion yields for Al and O as a function of $O_2$ exposure obtained for different primary ion energies: 0.65 keV (green asterisks), 1 keV (black squares), 1.5 keV (red circles) and 3 keV (blue triangles). Panel (a) and (c) show the ion signals for $A^+_{Al,AlO}$ and $A^+_{O,AlO}$ according to Eq. 3, respectively. In panel (b) the exposure curves are normalized to a constant initial ion yield of clean Al.

A comparison of the oxygen data sets shows, that the $O_2$ dose necessary to reach a steady equilibrium is constant for all investigated primary energies indicating a constant information depth. The corresponding intensities of the O signals show only a weak dependence on the energy, compared to the signal from the metal constituent. In fact, data are similar for the data sets with $E_0 \leq 1.5$ keV, whereas only at 3 keV the ion yield is significantly higher. Assuming equal surface composition for the same exposure dose, either the ion fraction for O scales weakly with energy in contrast to the $P^+_{Al}$ or information depth and ion fraction are changing such that the product only shows a weak energy dependency.

To summarize, in Fig. 2 and 3, two key features of the ion yield for Al and O, respectively, were observed. The reduction in the ion yield for Al in the transition from the clean to the oxidized state by a factor of ~ 4 can only partly be explained by different surface coverages. Consequently, the ion fraction, and therefore, the neutralization efficiency for Al has to be different for the clean metal and in the oxide. This matrix effect in the ion yield, however, is mostly independent of the primary energy as can be seen from the uniform exposure curves for Al in Fig. 3(b) resulting in an energy independent final exposure state. Alternatively, only a simultaneous change in information depth and ion fraction could explain this constant factor obtained for the different primary energies. For the O signal we observe similar ion yields at the final exposure dose for $E_0 \leq 1.5$ keV, only the 3 keV data deviate. A constant $A^+_{O,AlO}$ would require the product of surface coverage and ion fraction to be constant. With previous observations of only minor subsurface contributions in the ion yield for the present system and the assumption of the same surface coverage for a given exposure dose, the ion fraction has to have only a weak energy dependency in contrast to the typically studied metals. Note, the linear scale for the O signal in Fig. 3(c) vs. the logarithmic scale for the Al signal in Fig. 3(a). This observation is well in agreement with the analysis performed on the spectra shown in Fig. 2 requiring a strong difference in the energy scaling of the underlying neutralization mechanisms for O compared to Al.

To study the nature of the observed effects in more detail, and extract possible information on the charge exchange processes, we performed further measurements on the initial and final states of the exposure curves, without the constraint of continuous monitoring during exposure. Thus, in the following we compare for an increased number of energies the ion yield for Al and O in the clean and oxidized surface. To avoid any assumptions of the surface coverage in the oxide, Fig. 4 shows the normalized ion yield $A_i^+$, which is according to Eq. 3 proportional to the product of surface coverage and ion fraction, as a function of the inverse perpendicular velocity. The signals for a clean Al(111) (black squares) as well as the Al and O signal in the oxide are depicted as red points and blue triangles, respectively. Note, the single exponential scaling of the $A_i^+$ data sets as expected in the Auger regime according to Eq. 1, though for the investigated energies both Auger and resonant processes are expected to contribute to charge exchange. For the data for Al a characteristic velocity of $v_{c,Al}^{111} = (2.44 \times 10^5 \pm 1.25 \times 10^4)$ m/s was found (indicated as black solid line in Fig. 4), resulting in a comparable value (10% higher) than previously reported in literature [15]. The authors obtained the ion fraction with an ESA and ToF-LEIS system, where in the latter set-up both the yields of backscattered ions and neutrals are recorded. Ion fractions obtained in a ToF-system can therefore be evaluated independently of knowledge on the surface coverage.

The signals for Al obtained in the oxide are significantly lower than for the metallic surface with a scaling of $v_{c,Al}^{oxide} = (2.55 \times 10^5 \pm 1.05 \times 10^4)$ m/s (red solid line), resulting in the same energy dependency within the uncertainties between the ion fraction of Al in the metal and the oxide. The red dashed line in Fig. 4 considers the decrease in the number of Al atoms from a metallic Al(111) to a stoichiometric Al$_2$O$_3$ surface with the characteristic velocity of the metal $v_c^{111}$. Therefore, the large difference in the normalized yield between metal and oxide cannot exclusively be explained with a different surface coverage resulting in a neutralization efficiency depending on the chemical environment of the surface. Still, for the present system high surface sensitivity and the possibility of quantitative surface composition analysis is maintained as no energy dependence of the yield ratios is observed.

The normalized yields for O in Fig. 4 scale completely different with energy compared to Al or other metals with a characteristic velocity of $v_{c,O} = (4.4 \times 10^4 \pm 8.2 \times 10^3)$ m/s. The open triangles correspond to the O signals obtained from the detailed exposure curves in Fig. 3(b) at the steady equilibrium. The higher number of data points for the yield from O in the fully oxidized surface also confirms the weak energy dependence expected from the exposure curves. Within statistics, the dependence on the inverse perpendicular velocity follows the typical scaling with constant $v_c$. At energies < 0.8 keV, the neutralization efficiency of O is smaller compared to clean Al(111). In general, the weak energy dependency of $A^+_{O,AlO}$ indicates similar neutralization mechanisms in the whole investigated energy regime.

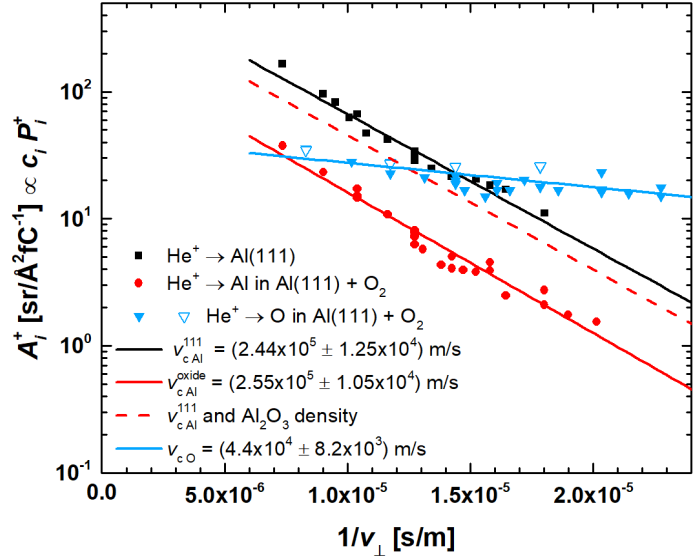

Fig. 4: Ion signals for He$^+$ scattered from Al in metallic (black squares) and oxidized surfaces (red circles) as well as O (blue filled and open triangles). Single exponential fits for the characteristic velocities are plotted for Al in the metallic (black solid line) [15] and oxidized surface (red solid line), as well as for O (blue solid line). The red dashed line indicates a normalized yield for Al under the assumption of a stoichiometric Al$_2$O$_3$, if the ion yield does not underlie any matrix effects.

With ongoing exposure, also differences in the peak positions are observed. In Fig. 5 the shift of the peak positions of Al (full symbols) and O (open symbols) as a function of O$_2$ exposure is plotted for the different primary energies. The 3 keV data set has been corrected for a constant drift in the current measurement. For all investigated energies, the peak position of Al shifts by ~ 5 eV to ~ 8 eV towards lower final energies during surface oxidation. The corresponding evaluation of the peak positions of O yields either no change or an increase up to ~ 3 eV. These opposing trends cannot be explained with charging effects of the surface. A location of the O-atoms in front of the outermost Al layer, however, as obtained in several different studies [24–27] would result in an additional energy loss of the projectiles scattered from Al when traversing the O layer, whereas the peak position of O would remain mostly unaffected.

The inset in Fig. 5 shows that during exposure to O$_2$ the FWHM of the Al peak remains constant. However, as indicated in Fig. 1, we cannot disentangle contributions from the surface for projectiles surviving AN or being re-ionized. Therefore, we cannot draw any conclusions concerning Auger and resonant neutralization efficiencies separately, but we observe that any changes of both neutralization mechanisms are such, that the width of the surface peak stays constant.

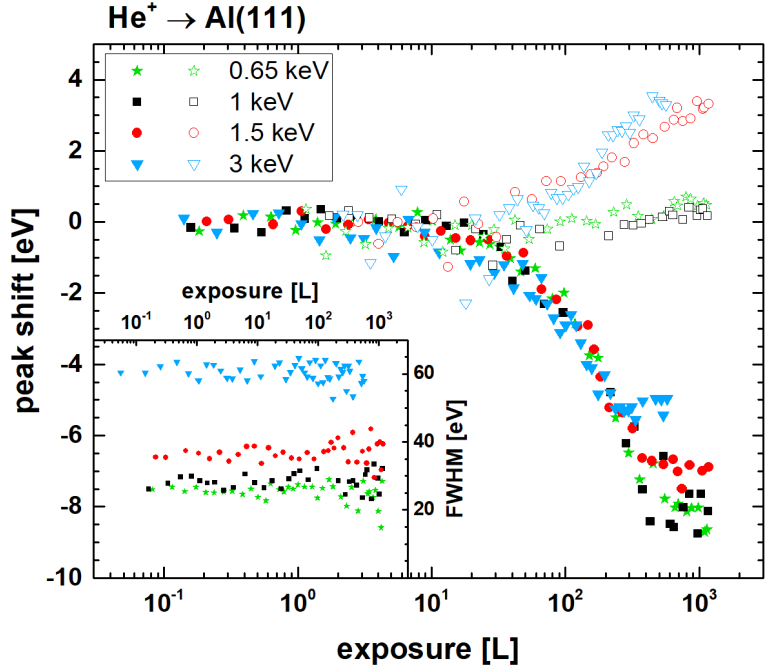

Fig. 5: The main plot shows the shift of the peak position of Al (full symbols) and O (open symbols) with $O_2$ exposure, whereas the inset depicts the FWHM of the Al signal. Both panels show data sets obtained for different primary energies: 0.65 keV (green asterisks), 1 keV (black squares), 1.5 keV (red circles) and 3 keV (blue triangles).

3.3 Ta(111) surface

The ion yields of $He^+$ scattered from a Ta(111) surface as a function of $O_2$ exposure are plotted in Fig. 6 for both (a) Ta and (b) O signals. The exposure has been performed for different primary energies, i.e., 0.85 keV (green asterisks), 1.5 keV (black squares), 2 keV (red circles) and 3 keV (blue triangles). Panel (a) shows the exposure curves for Ta which are normalized to the initial value of clean Ta to improve the comparability between the data sets for different primary energies – analog to Fig. 3(b) for Al. Already at sub-Langmuir exposures a decrease in the Ta signal is observed and altogether only several tens of Langmuir are necessary to approach an equilibrium state of the ion yields for Ta and O, in accordance with literature were exposure doses of ~ 10 L are reported [22]. Note, in this context, due to the very open structure of the bcc (111) surfaces, a larger contribution from sub-surface layers can be expected. The fact that the oxidation process for Ta starts at lower $O_2$ doses compared to Al indicates a higher initial sticking probability for O on clean Ta. The curvatures of the exposure data for Ta, however, are for all primary energies lower than for the oxidation process of Al. The oxygen doses necessary until a stable state is reached, are furthermore depending on the employed primary energy. We show two different fits according to the same statistical growth model as used for Al for the data sets with the lowest and highest investigated primary energy. For 0.85 keV good agreement between the experimental data and the fit (green dashed line) is found with the final oxidation stage reached at ~ 10 L in accordance to [22]. However, for the 3 keV data a single exponential fit (blue line) requires an about twice as large parameter $L_0$ to sufficiently describe the oxidation process. This difference between the exposure curves can be understood in terms of the open structure of the Ta(111) surface where sub-surface signals are expected to contribute to the ion yield for Ta resulting in different information depths depending on the primary energy. Therefore, the 0.85 keV data set with the lowest information depth reaches a saturation already at ~ 10 L indicating a completely oxidized first monolayer. For the 3 keV data, sub-surface layers contribute stronger to the ion signal and therefore, sub-surface oxidation continuous to influence the ion yield for Ta resulting in a slower decrease of the ion signal and a higher oxygen dose necessary to reach a final equilibrium. Also the intensity of the ion yields of Ta at the final oxygen coverage differs between the investigated primary energies, which is in contrast to the observations for Al(111) featuring a closed packed surface.

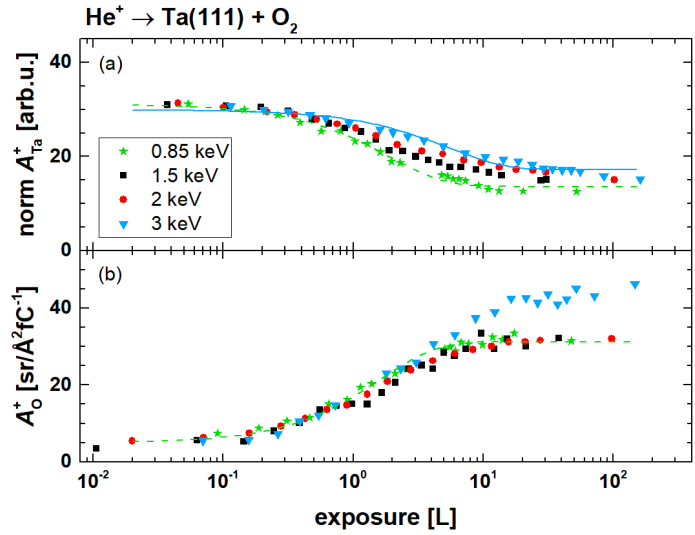

Fig. 6: Ion yields of He$^+$ scattered from (a) Ta and (b) O as a function of O$_2$ exposure. All Ta signals are normalized to one initial ion yield before the exposure, whereas the ion yields for O are normalized according to Eq. 3. The exposure has been performed for primary energies of 0.85 keV (green asterisks), 1.5 keV (black squares), 2 keV (red circles) and 3 keV (blue triangles).

In panel (b) the corresponding ion signals $A^+_{O,TaO}$ are plotted. Note, for all data sets the saturation of the O signal is observed at the same O$_2$ dose with similar curvatures, in contrast to the corresponding metal data. The yield for the oxidized surface shows only a weak dependence on $E$ with a deviation for the highest energy. These observations indicate a less pronounced contribution from sub-surface layers, i.e. different information depth for the oxygen signal as compared to Ta. Again, the data also indicate a weak energy dependence of the ion fraction for O as it was observed in Al.

To gain further insight in the neutralization efficiencies for Ta and O we perform a similar analysis of either the metallic or oxidized surface for a series of primary energies – analog to Al. In Fig. 7 normalized yields for Ta in the metallic and oxidized surface and O as a function of the inverse perpendicular velocity are depicted. Similar to Al, the yields can be described with a single exponential function, though both Auger and resonant processes can occur in the entire investigated energy regime. The black squares and red circles correspond to normalized ion yields obtained for Ta in a metallic and oxidized surface, respectively, whereas the signals for O are depicted as blue triangles.

An exponential fit for the ion yield of Ta in the metal results in a characteristic velocity of $v_{c,Ta}^{111} = (2.25\times10^5 \pm 9.6\times10^3)$ m/s, depicted as black solid line in Fig. 7. According to the classification of Rusch et al., [29] we do not observe oscillations in the ion yield of Ta independent of the surface structure within the experimental uncertainties. This finding is in contrast to experimental results from Arikawa et al., [30], who observed small oscillations in the ion yield for He$^+$ scattered from polycrystalline Ta. The oscillations with different periodicity were attributed to interactions of the He ion with the 4f$_{5/2}$ and 4f$_{7/2}$ electrons. Our obtained $v_{c,Ta}^{111}$ is smaller compared to the characteristic velocity from Al(111) with $v_{c,Al}^{111} = (2.44\times10^5 \pm 1.25\times10^4)$ m/s resulting in a weaker energy dependency of the neutralization efficiencies. The extrapolation of the clean Ta signal to $1/v_c \to 0$ is lower compared to Al by a factor of ~ 5.5. The lower atomic density in a bcc (111) monolayer compared to an fcc (111) layer by a factor of ~ 2.6 can only partly describe the observed difference, indicating higher neutralization efficiencies in Ta compared to Al. Assuming a stoichiometric Ta$_2$O$_5$ surface and the characteristic velocity of Ta(111), i.e. no matrix effect in the ion yield, the data points obtained in the oxide would need to coincide with the data from the clean metal, indicated as the red dashed line in Fig. 7. However, a single exponential fit to the experimental data yields a characteristic velocity of $v_{c,Ta}^{oxide} = (3.13\times10^5 \pm 1.2\times10^4)$ m/s. An extrapolation of the Ta signal for both surfaces towards $1/v_c \to 0$ yields the same offset within the uncertainties corresponding to a Ta-atom density as in a stoichiometric Ta$_2$O$_5$. The completely different scaling of the ion yield with energy between the metal and the oxide can only be explained with neutralization efficiencies depending on both energy and surface composition. In comparison, for Al we observed a non-energy-dependent matrix effect of the ion yield.

The energy dependency of the O signal is weaker as for Ta with a characteristic velocity of $v_{c,Ta} = (2.5\times10^4 \pm 1.3\times10^4)$ m/s, indicating a different neutralization behavior of O compared to Ta. This result is in accordance to the observations for $A_{O,AlO}^+$ where the energy scaling of the O signal is also significantly weaker than for Al, though with a higher $v_c$ for the $A_{O,AlO}^+$ data. Note, the open triangles in Fig. 7 represent the O signals from the saturation point in the exposure curves where the intensity of the ion yield was found to be constant for $E_0 \leq 2$ keV. For all investigated energies in Fig. 7, it can be observed that $A_{O,TaO}^+ > A_{Ta,Ta}^+$. Due to the proportionality of the plotted signals to the product of surface coverage and ion fraction, either a significant higher O-atom density or a lower neutralization efficiency for O compared to Ta is necessary. However, even with a factor of 2 in the surface atom densities the neutralization efficiency for Ta would be higher for energies below 2 keV.

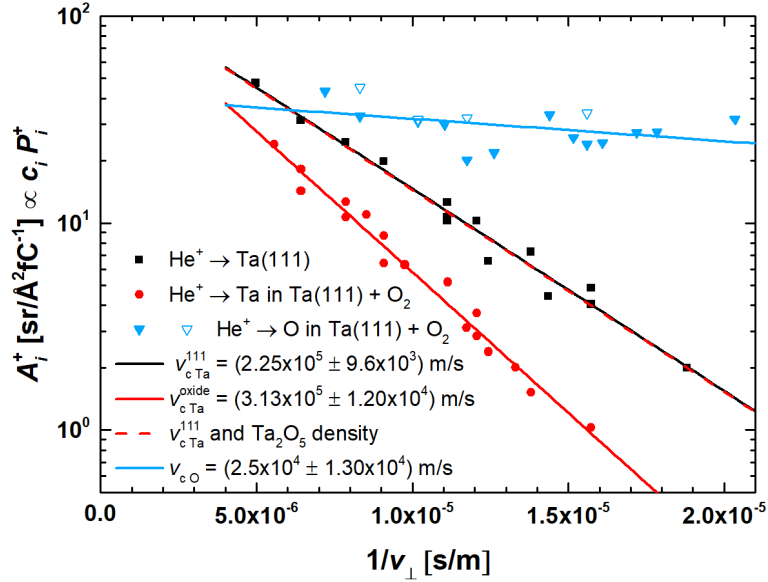

Fig. 7: Normalized He$^+$ ion yield (according to Eq. 3) as a function of the inverse perpendicular velocity. The black squares and the red circles correspond to data for scattering from Ta in the metallic and oxidized surfaces, respectively. The yields for O are depicted as blue triangles. The solid lines in the corresponding colors represent single exponential fits to the experimental data. The red dashed line indicates the ion yields for Ta in the oxide, if no matrix effects occur.

The evaluation of the peak position of Ta and O in the metal and oxide shows a similar behavior as for Al: the Ta peak shifts towards lower energies by ~ 5 eV to ~ 8 eV, whereas the position of the O peak can be found to increases by up to ~ 2 eV. Similarly as for Al, this opposing trend excludes any charging effects of the surface and might indicate an O layer in front of the first Ta layer. The FWHM of the Ta signal is independent of the O$_2$ exposure – similarly as for Al, which does not provide additional information on the neutralization efficiencies.

## 4. Summary & Conclusions

We measured the ion yields for He$^+$ backscattered from single crystalline Al(111) and Ta(111) as a function of O$_2$ exposure for four different primary energies. For Al(111) the ratio of the ion yields of the metal for the clean and oxidized surface is independent of the primary energy. Based on the observation from previous investigations that the ion yield only shows minor sub-surface contributions for less open surfaces, this constant ratio implies that the ion fraction of Al in the pure metal and its oxide exhibits the same energy dependency. The absolute values of the ion fractions, however, are found different from the expected surface structures, which constitutes an energy-independent matrix effect.

For the Ta(111) surface, dependent on the employed primary ion energy, different oxygen doses are necessary to reach saturation in the ion yields for Ta in the oxidation process. This observation indicates different information depths, attributed to the open bcc (111) structure. Additionally, the ion yields for Ta scale different with energy in metal and oxide resulting in neutralization efficiencies dependent on both, energy and surface composition.

For the oxidation of both Al and Ta, the $O_2$ dose necessary to reach a saturation in the ion yield for O was found to be independent of the primary energy. This fact indicates similar information depths for O, i.e. high surface sensitivity in both systems. Compared to the metal constituents and the majority of systems described in literature [11] only a weak energy dependence of the O signal in the oxidized samples is observed. Within experimental uncertainty, the O signal does not show a dependence on the matrix, a result substantially different from what has been observed for other light species e.g. carbon [31]. At the same time, this observation is in accordance with expectations of the electronic configuration of O in different oxides being similar with a strongly covalent character of the bond with all valence electrons located at the anion and identical interaction with slow ions [32].

The peak position for both metals shifts towards lower energies with ongoing exposure, whereas the position of the O peak does either not change at all or only slightly increases. These opposing effects allow the exclusion of any charging effects and point towards a preferential location of the O atoms in front of the first metal layer.

**Acknowledgment**

The authors acknowledge the help of Simon Schuler in collecting the data. We acknowledge financial support by the Austrian Science Fund FWF within project P25704 (NEO-LEIS). Additionally, BB is grateful to the Wilhelm-Macke Foundation at the JKU for partial support of her stay at UU.